\documentclass[aip,twocolumn,letterpaper]{revtex4}
\pdfoutput=1
\def\b#1{\mathbf{#1}}
\def\h#1{\hat{#1}}

\def\p{\partial}

\def\ss{\color{red} \bigodot \color{black}}

\def\ss#1{\mbox{sin}^{2}{#1}\;}

\usepackage{caption}
\usepackage{fullpage}
\usepackage{graphics}
\usepackage{epsfig}
\usepackage{xcolor}
\usepackage{amsmath}
\usepackage{amssymb}
\usepackage{subfig}

\begin{document}
\title{Magnetic nulls in interacting dipolar fields}
\author{Todd Elder and Allen H. Boozer}
\affiliation{Columbia University, New York, NY  10027\\ tme2123@columbia.edu \textnormal{and} ahb17@columbia.edu}

\begin{abstract}
	The prominence of nulls in reconnection theory is due to the expected singular current density and the indeterminacy of field-lines at a magnetic null. Electron inertia changes the implications of both features. Magnetic field lines are distinguishable only when their distance of closest approach exceeds a distance $\Delta_d$.  Electron inertia ensures $\Delta_d\gtrsim c/\omega_{pe}$. The lines that lie within a magnetic flux tube of radius $\Delta_d$ at the place where the field strength $B$ is strongest are fundamentally indistinguishable. If the tube, somewhere along its length, encloses a point where $B=0$,vanishes, then distinguishable lines come no closer to the null than $\approx (a^2c/\omega_{pe})^{1/3}$, where $a$ is a characteristic spatial scale of the magnetic field. The behavior of the magnetic field lines in the presence of nulls is studied for a dipole embedded in a spatially constant magnetic field.  In addition to the implications of distinguishability, a constraint on the current density at a null is obtained, and the time required for thin current sheets to arise is derived. 
\end{abstract}

\date{\today} 
\maketitle

\section{Introduction \label{sec:Intro}}
\subsection{Magnetic reconnection \label{sec:Intro-MR}}

Magnetic nulls have long been prominent in the theory of magnetic reconnection. The traditional reasons for their importance are the  indeterminacy of field lines at a null and the formation of a singular current density along field lines that intercept a null \cite{Craig:Singlar_current}.

Magnetic field lines have their closest approach at the thinnest locations of flux tubes and can be separated by great distances at other places along the lines. Resistivity or electron inertia need only make magnetic field lines indistinguishable across the thinnest parts of flux tubes to produce reconnection \cite{Boozer:MR+Thermal_equilibration}. Field line indistinguishability is equivalent to the breaking of the frozen in condition of field lines.

Recognition of the importance of locations of the closet approach of magnetic field lines is particularly great for understanding the role of nulls. Electron inertia sets a minimum distance over which magnetic field lines are distinguishable, $\Delta_d \gtrsim c/\omega_{pe}$; plasma resistivity can make $\Delta_d$ even larger. Electron inertia produces magnetic flux tubes of indistinguishable lines when the tube has a radius of $\Delta_d$ anywhere along its length.  Since magnetic flux is constant along a magnetic flux tube, the spatial extent of a magnetic flux tube that passes near a null becomes spatially extended.  Regions of field-line indistinguishability are not determined by regions near nulls but essentially where the field strength $B$ has a maximum along each tube. 

Magnetic nulls are points where magnetic field lines of different topologies come arbitrarily close to one another, which can and will be studied by placing a sphere of infinitesimal radius about each null.  But, the regions over which magnetic field lines are fundamentally indistinguishable in an evolution is determined by the $\Delta_d$ scale compared to the size of flux tubes far from the nulls.

The subtlety of magnetic nulls, the $\Delta_d$ scale of indistinguishability, and topology will be studied in what may be the simplest well-defined, but common situation, the interaction of a magnetic dipole with a constant magnetic field.  Dipoles have a singularity, but in nature dipole fields, such as the dipole field of the earth, are enclosed within what is essentially a spherical object.  This will be assumed, and the topological properties of the magnetic field will be studied by field line mappings between the sphere that encloses the dipole, infinitesimal spheres around the nulls, and a distant sphere on which the uniform magnetic field is dominant. 

A caveat on terminology: in astrophysics, a magnetic flux tube can imply that the magnetic field is far stronger within the tube than without.  Magnetic flux tubes are too important for understanding magnetic fields that vary smoothly in space to allow this implication to impede thought.  Here a magnetic flux tube is defined by a region enclosed by the magnetic field lines that pass through a closed curve, such as a circle.

%

\subsection{Brief history of nulls \label{sec:Intro-History}}
The study of magnetic nulls has historically been motivated by the study of Earth's magnetosphere. Magnetic reconnection at magnetic nulls was posited as the driving force for convection in Earth's magnetosphere \cite{Dungey-1961}. 

In 1963, Dungey \cite{Dungey-1963} proposed a simple model for the formation of magnetic structures of vanishing strength: that of a magnetic dipole immersed in a uniform field antiparallel to the magnetic dipole moment. This symmetric configuration produces a line of vanishing magnetic field strength, or a line null. 

In 1973, Cowley \cite{Cowley-1973} considered the effect of a non-symmetric configuration: a magnetic dipole immersed in a uniform magnetic field oriented at an angle $\zeta$ with respect to the magnetic dipole moment. This configuration produces two magnetic point nulls: three-dimensional points at which the magnetic field strength vanishes, and illustrates well the general principle that a line null breaks into well-separated point nulls in the presence of an arbitrarily small perturbation. Cowley's work formed the underpinnings of null research such as the spine-fan structure of the magnetic field near a null. Stern \cite{Stern-1973} also independently considered the effect of non-symmetric configurations in 1972.  

The work of Dungey, Cowley, and Stern have formed the basis for magnetic null research to the present day. More recent considerations include the motion of null points \cite{Muphy:motion-of-nulls}, the development of analytic solutions of steady-state reconnection \cite{Craig:Exact-solns-nulls}\cite{Galsgaard:Exact-solns-nulls}, and their role in particle acceleration \cite{Dalla:Nulls-and-particle-acceleration}\cite{Boozer:Particle_Acceleration}. 
	
Nulls have been observed both in Earth's magnetosphere \cite{Xiao:Nulls-in-magnetosphere} and in the solar corona \cite{OFlann:Nulls-in-corona}.

We will use a simplified model of Earth's magnetosphere interacting with the solar wind in order to study properties of magnetic nulls.

\subsection{Magnetic nulls \label{sec:Intro-Nulls}}
In this section we describe well-known results of nulls, \cite{Dungey-1963} \cite{Cowley-1973}  \cite{Parnell-1996}.

Magnetic nulls are three-dimensional points where the magnetic field strength vanishes. 
A first-order Taylor expansion of a general divergence-free field around a point null has the form,
\cite{Parnell-1996}\cite{Boozer:MR-with-null+X-points}:
\begin{equation}
\b{B}(\b{x}) = \tensor{M}\cdot\b{x} + \dfrac{\mu_0}{2}\b{j}_0\times\b{x}
\label{eqn:Null-field}
\end{equation}
where
\begin{equation}
\tensor{M} = \dfrac{B_n}{2a}
\begin{bmatrix}
(Q_n+1) &     0    &  0\\
0    & -(Q_n-1) &  0\\
0    &     0    & -2\\
\end{bmatrix}
\end{equation}
When $a$ is the radius of the dipole sphere $B_n$ has a value equal to the typical field strength on that sphere.The three diagonal elements are dimensionless numbers such that the trace must is zero since the magnetic field is divergence free. $Q_n$ is a coefficient describing the topological structure of the null spine-fan. $\b{j_0}$ is the current density at the null. Section \ref{sec:F-L_Properties-Current_at_null} will show that the Lorentz force associated with the current $\mathbf{j}_0$ has a curl when $|Q_n|$ is not equal to unity, which means the force cannot be balance by the gradient of a scalar pressure.

Magnetic fields given in equation (\ref{eqn:Null-field}) with $\b{j}=0$ form the typical spine-fan structure of a null; see Fig. \ref{fig:Null-Sphere}. The spine-fan structure is either of type A: field lines coming in along the spine, a cylindrical column of magnetic field lines, and out along the fan, a plane of field lines, or type B: in along the fan and out along the spine.

When $\b{j}_0=0$, $Q_n$ describes the flow strength of field-lines in the fan plane. As $Q_n\to 0$, field-lines travel at an equal rate in both directions of the fan plane. When $Q_n\to 1$, field-lines travel increasingly unidirectionally in the fan plane. For $Q_n=1$, the fan of  collapses into a column of flux, forming a ``spine-spine" structure. $Q_n=1$ occurs when when the uniform field points only in the $-\h{z}$ direction. In this case, a line null forms around the equator of the dipole sphere with radius $a_N = \sqrt[3]{\mu/B_c}$. 
An overall multiplying factor can be removed and placed in front of the matrix.  The largest coefficient in absolute value can be set equal to -2 by an appropriate choice of the constant in front of the matrix and the coordinates can be chosen so that is the z-directed component.  The other two components must add up to +2.  This can always be achieved by making the larger of the two equal $Q_n+1$ and make it the x-directed component with $1\geq Q_n \geq 0$ and the other component, which must be y-directed, equal to $2 - (Q_n+1)$.



In ideal theory, a singular current density with a finite total current will generically arise there during an evolution \cite{Craig:Singlar_current}.  But, as noted in  \cite{Boozer:Less_current_MR}, it is difficult to concentrate a current in a narrow region across indistinguishable magnetic field lines as would be required to have an singular current density with a finite total current. Section \ref{eqn:Alfven_transit_time} shows that a thin current sheet can develop along field lines that pass near the nulls, but a long time is required for it to develop.


\subsection{Features of magnetic nulls in interacting dipolar fields \label{sec:Intro-Our_work}}
Sections \ref{sec:Model} and \ref{sec:F-L_Properties} examine Cowley's orginal model through the lens of surface mappings and magnetic flux tubes. A thought experiment involving electron inertia effects is used to describe a fundamental features missed in the literature.

\section{Model \label{sec:Model}}
\begin{figure}[h!]
	\vspace{0cm}
	\subfloat[][]{\includegraphics[height=150pt]{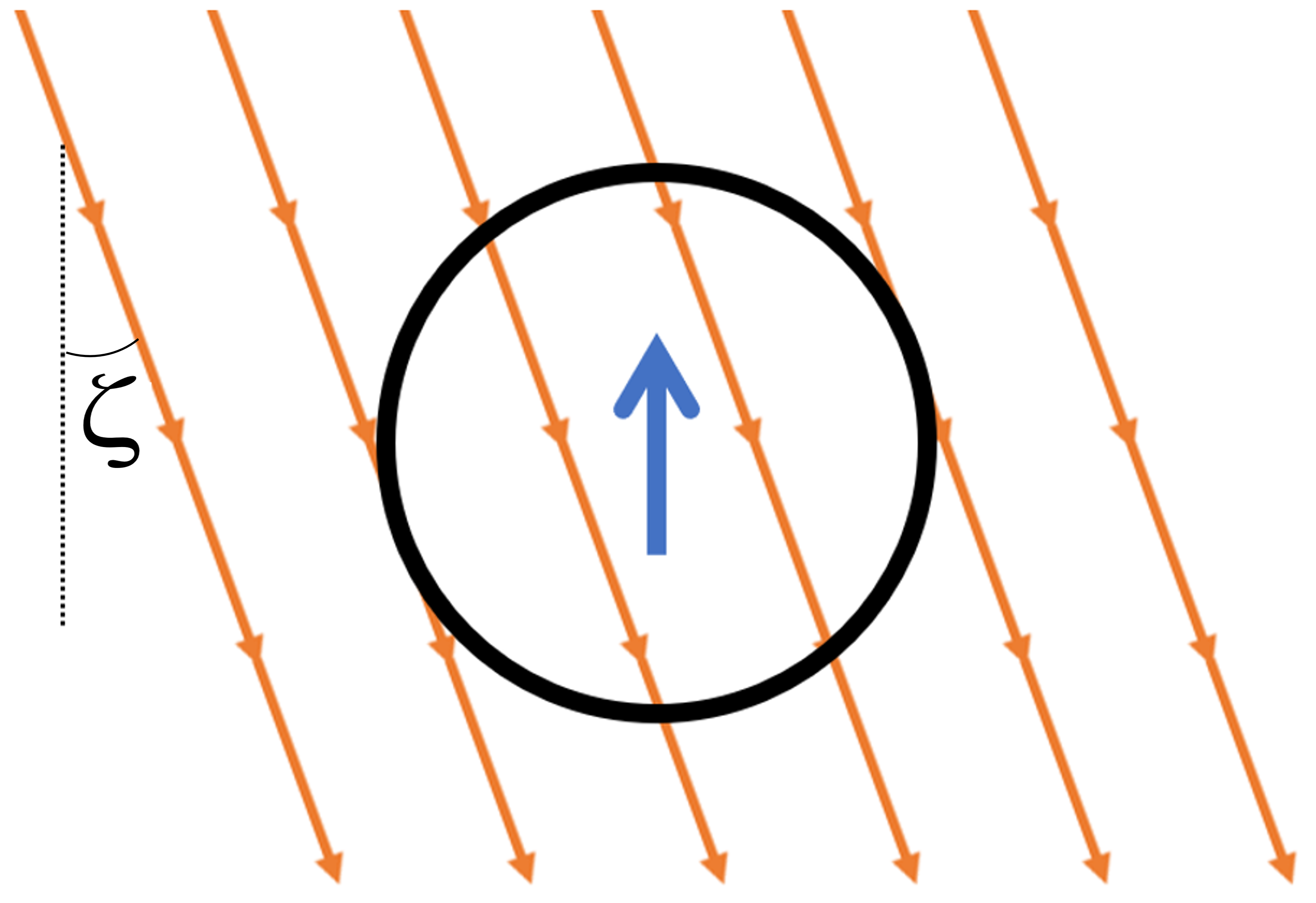}}
	\caption{Magnetic dipole (blue arrow) embedded in a spherical shell (black line) placed in uniform background magnetic field oriented at angle $\zeta$ (orange lines)}
	\subfloat[][$\zeta=0$]{\includegraphics[height=120pt]{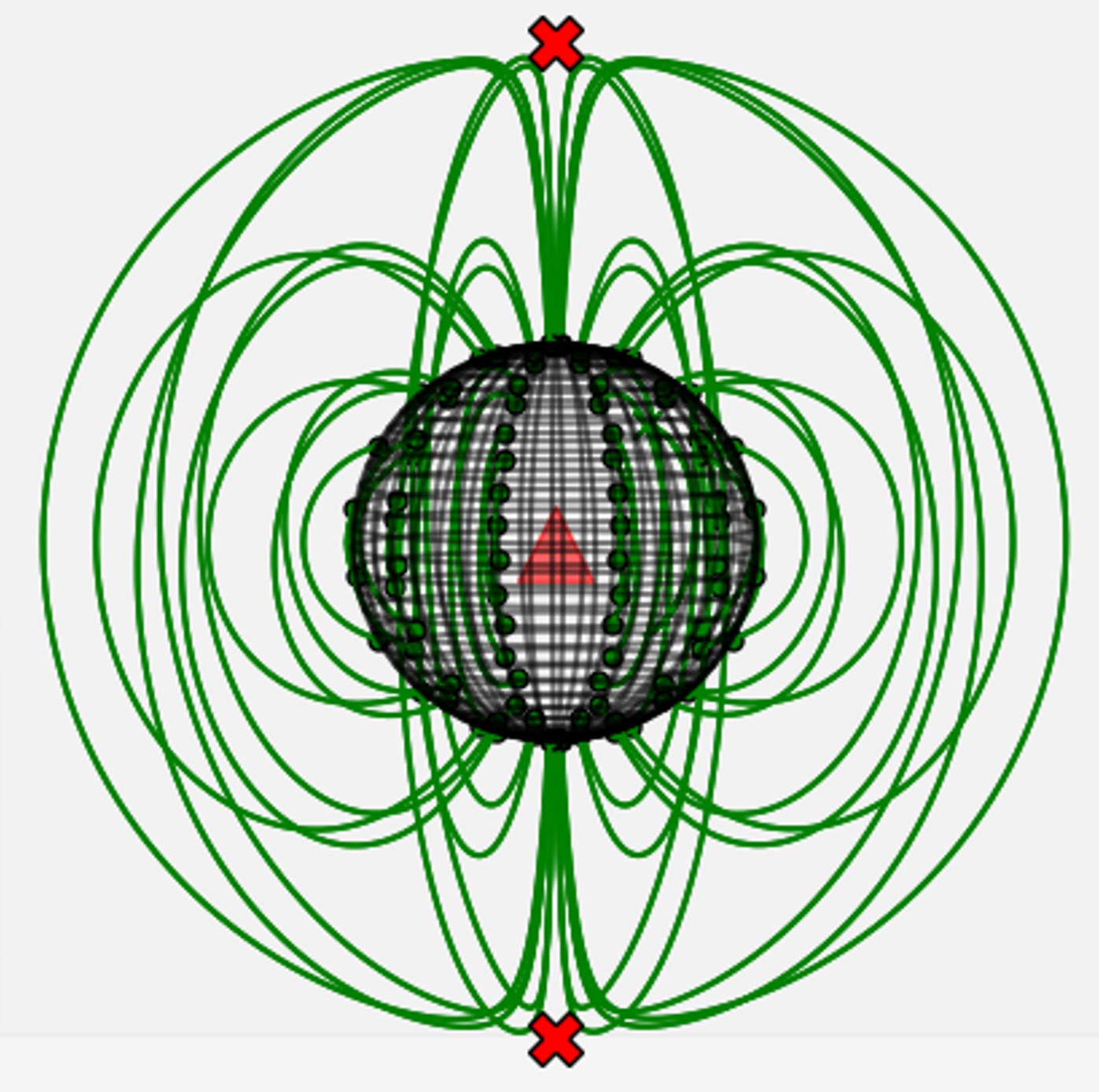}}
	\hspace*{0.1cm}
	\subfloat[][$\zeta=\pi$]{\includegraphics[height=120pt]{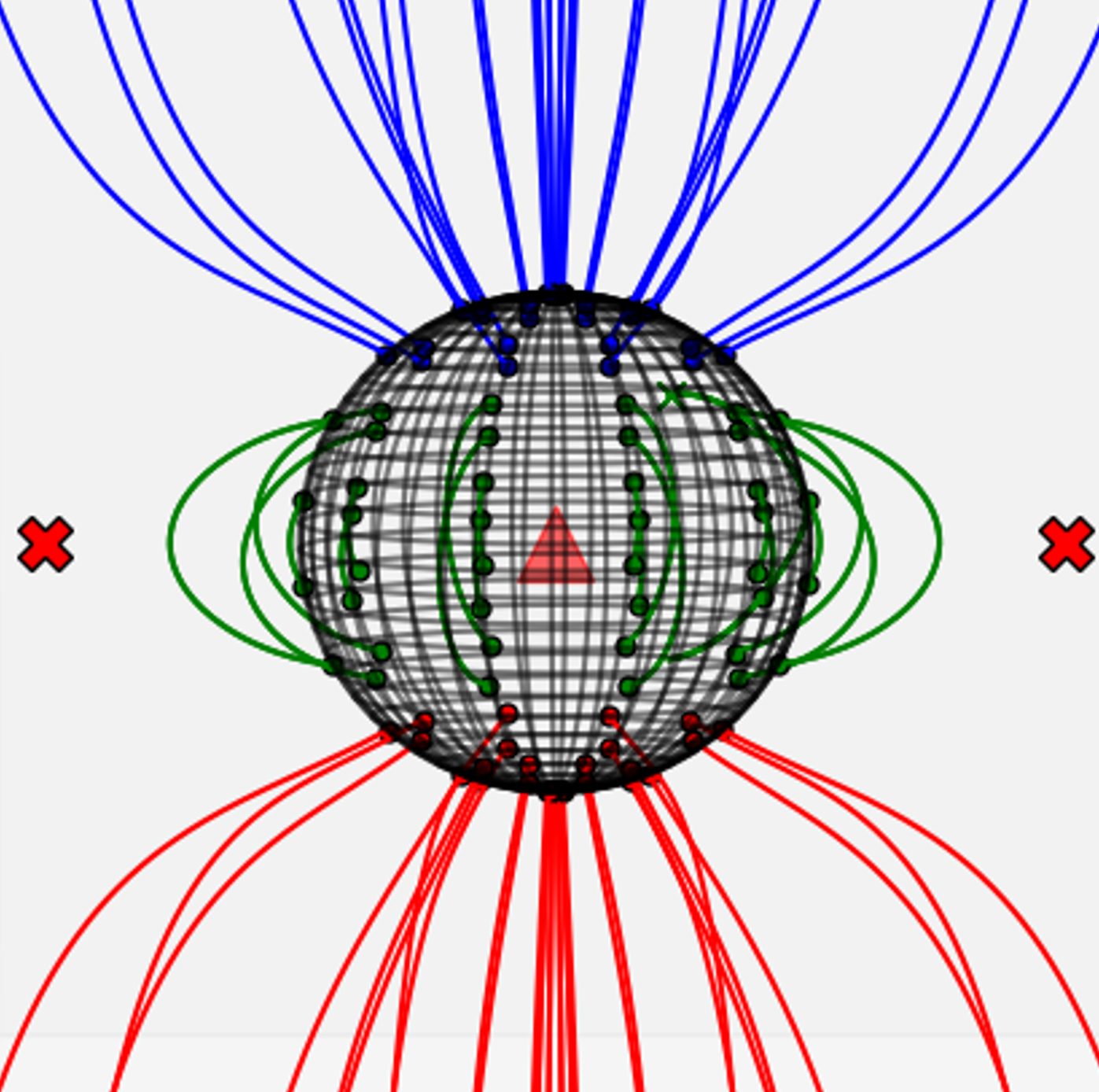}}
	\caption{Example fields for given $\zeta$. Lines shown are magnetic field lines; different colors denote topological type as described in Fig. \ref{fig:Mag-surf}. The sphere is the dipole sphere, the red X's are the locations of the magnetic nulls. Note that for $\zeta=\pi$ there is actually an equatorial line null connecting the two point nulls shown}
	\label{fig:Model}
\end{figure}

In nature, magnetic dipoles arise embedded in some medium. An example is the Earth, which can be taken to be a magnetic dipole embedded in a sphere. A purely dipolar magnetic field has a singularity at the location of the dipole, which when included makes field line descriptions confusing. Thus, we consider a magnetic dipole embedded in a sphere of radius $a$ placed into a uniform background magnetic field oriented at an angle $\zeta$; see Fig.\ref{fig:Model}. The theory is of general importance in magnetic fields that have nulls.   One example is a simplified static model of the interactions between the Earth's magnetic field and the solar wind.

The sphere serves to (i) provide a distinction between field lines of different topological types on the surface of the dipole sphere and (ii) eliminate the magnetic singularity located at the dipole moment. Field lines are not tracked below the surface of the dipole sphere.

For a dipole with a magnetic moment $\mathbf{m}_d=m_d\hat{z}$, the magnetic field in spherical coordinates is 
\begin{eqnarray}
\b{B}_d &=&  \frac{\mu}{r^3}(2\cos\theta \hat{r} + \sin\theta \hat{\theta})
\end{eqnarray}
where $\mu \equiv (\mu_0/4\pi)m_d$. The constant field is $\mathbf{B}_c=B_c(\cos\zeta\hat{z}+\sin\zeta\hat{x})$ or in spherical coordinates,  
\begin{eqnarray}
\b{B}_c & =& B_c (\cos\zeta\cos\theta+ \sin\zeta\sin\theta \cos\varphi ) \hat{r} \nonumber \\
&& - B_c (\cos\zeta\sin\theta  - \sin\zeta\cos\theta \cos\varphi) \hat{\theta} \nonumber \\
&& -B_c \sin\zeta\sin\varphi \;\hat{\varphi}. 
\end{eqnarray}
Nulls form where $\b{B}_d=-\b{B}_c$.

For reference, the cartesian-spherical transformation equations are
\begin{eqnarray}
x &=& r\sin\theta \cos\varphi \notag\\
y &=& r\sin\theta \sin\varphi \label{eqn:Cart_Sph_transforms}\\
z &=& r\cos\theta \notag
\end{eqnarray}

Field-lines may have different topologies. Topological type is determined by the boundary conditions of a field-line integrated forwards and backwards. In this system there are four topological types. Field lines: (i) start and end on the surface of the sphere, (ii) start on the sphere and end up infinitely far away, (iii) begin infinitely far away and land on the sphere, (iv) never touch the surface of the sphere; see Fig. \ref{fig:Mag-surf}a.

This idea of special field-line mappings associated with the magnetic nulls is related to the idea of the magnetic skeleton \cite{Priest:Magnetic-skeleton}.


\subsection{Topological features \label{sec:F-L_Properties-Topology}}
\subsubsection{A magnetic surface \label{sec:F-L_Properties-Symmetry_plane}}

\begin{figure}[h!]
	\centering
	\vspace{0cm}
	\subfloat[][]{\includegraphics[height=200pt]{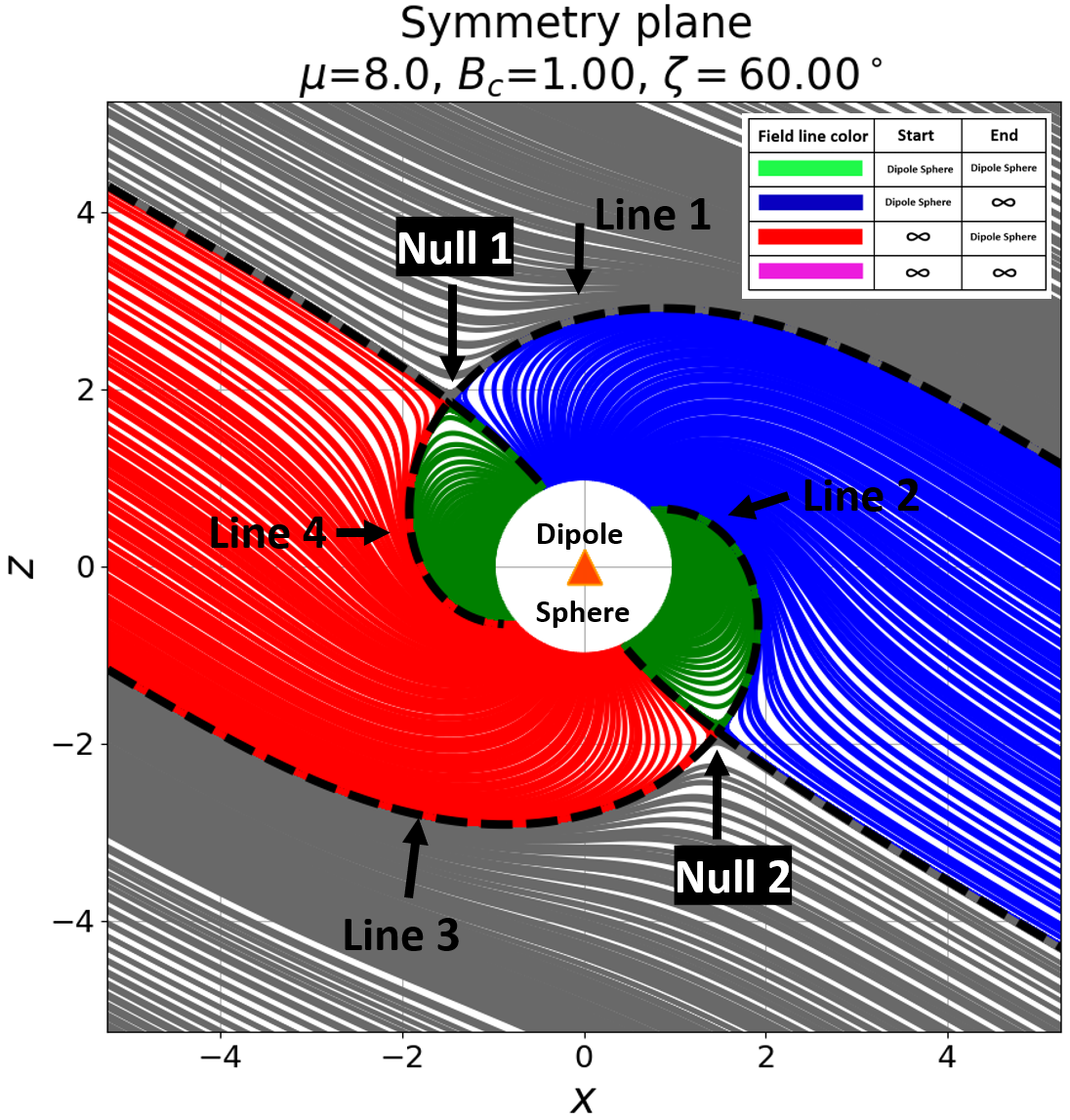}}
	\\
	\subfloat[][]{\includegraphics[height=100pt]{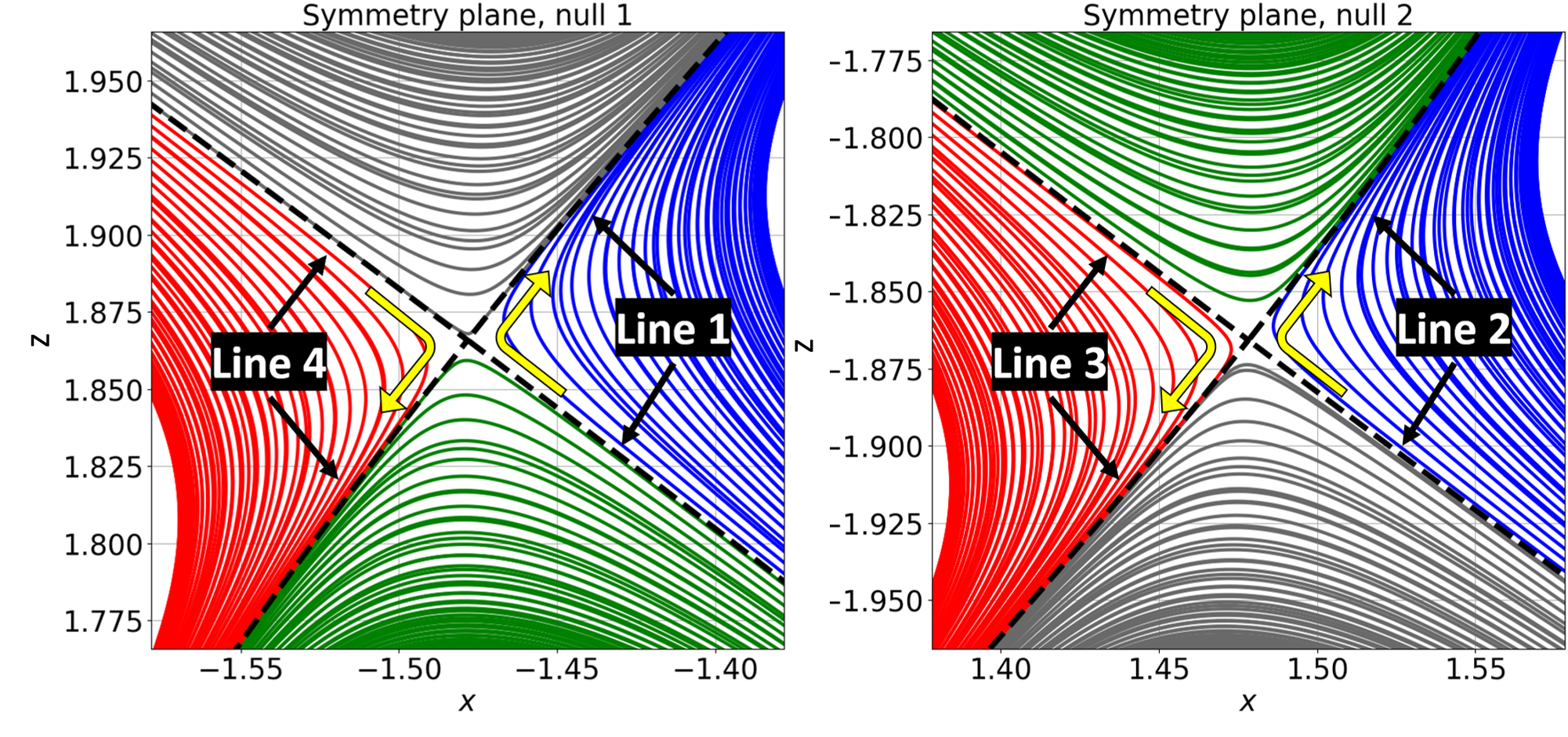}}
	\caption{Field-lines in the magnetic surface $y=0$. The magnetic nulls lie in this plane and accordingly so does each topological type. The characteristic field lines (lines 1,2,3,4) map out the separatrices of the system. In Figure b, the nulls are zoomed into to show the characteristic field-line paths near the null.}
	\label{fig:Mag-surf}
\end{figure}

The $y=0$ plane is a magnetic surface since the magnetic field normal to that surface, $B_{\varphi}$, is zero. Any field-line which starts in the $y=0$ plane remains in it. This plane provides a convenient description of the entire system which requires only two dimensions, See Fig. \ref{fig:Mag-surf}.  The two point-nulls that arise when a dipole lies in a uniform field lie in this plane.

Within the magnetic surface, four field-lines characterize the topological properties of the entire 3D system, Fig.\ref{fig:Mag-surf}a.  Lines 1 and 3 comprise the null spine as they travel out from the dipole sphere to nulls 1 and 2, then become fan lines as they pass through the null. These lines outline a separatrix in the field far away from the null which distinguishes $\infty$,$-\infty$ and $\infty$-surface mappings. Lines 2 and 4 go into the null fan as they travel out from the dipole sphere to the nulls and then become the null spine in the far-field and also act as a separatrix between $\infty$,$-\infty$ and $\infty$-surface mappings. 

As we shall see in Sec. \ref{sec:F-L_Properties-Topology_4sphere} figures \ref{fig:4-spheres}abc, these four field lines on the magnetic surface provide a characteristic description of the separatrices arising in this system.

\subsubsection{Mappings on four different spheres \label{sec:F-L_Properties-Topology_4sphere}}
The system may be considered a mapping between four spherical surfaces: (i) the dipole sphere of radius $a_D$, (ii)+(iii) two 
spheres of radius $a_N<<a_D$ placed about each null, and (iv) a spherical shell centered about the magnetic dipole with radius $a_\infty>>a_D$. For our study,  $a_\infty>>a_D>>a_N$. $a_\infty$ is large enough that $\b{B}(a_\infty,\theta,\phi)\approx \b{B}_c(a_\infty,\theta,\phi)$.

\begin{figure}[h!]
	\centering
	\vspace{0cm}
	\subfloat[][Null Sphere. Each different topological type are present near the nulls. Along the black dotted line, $B_r=0$]{\includegraphics[height=125pt]{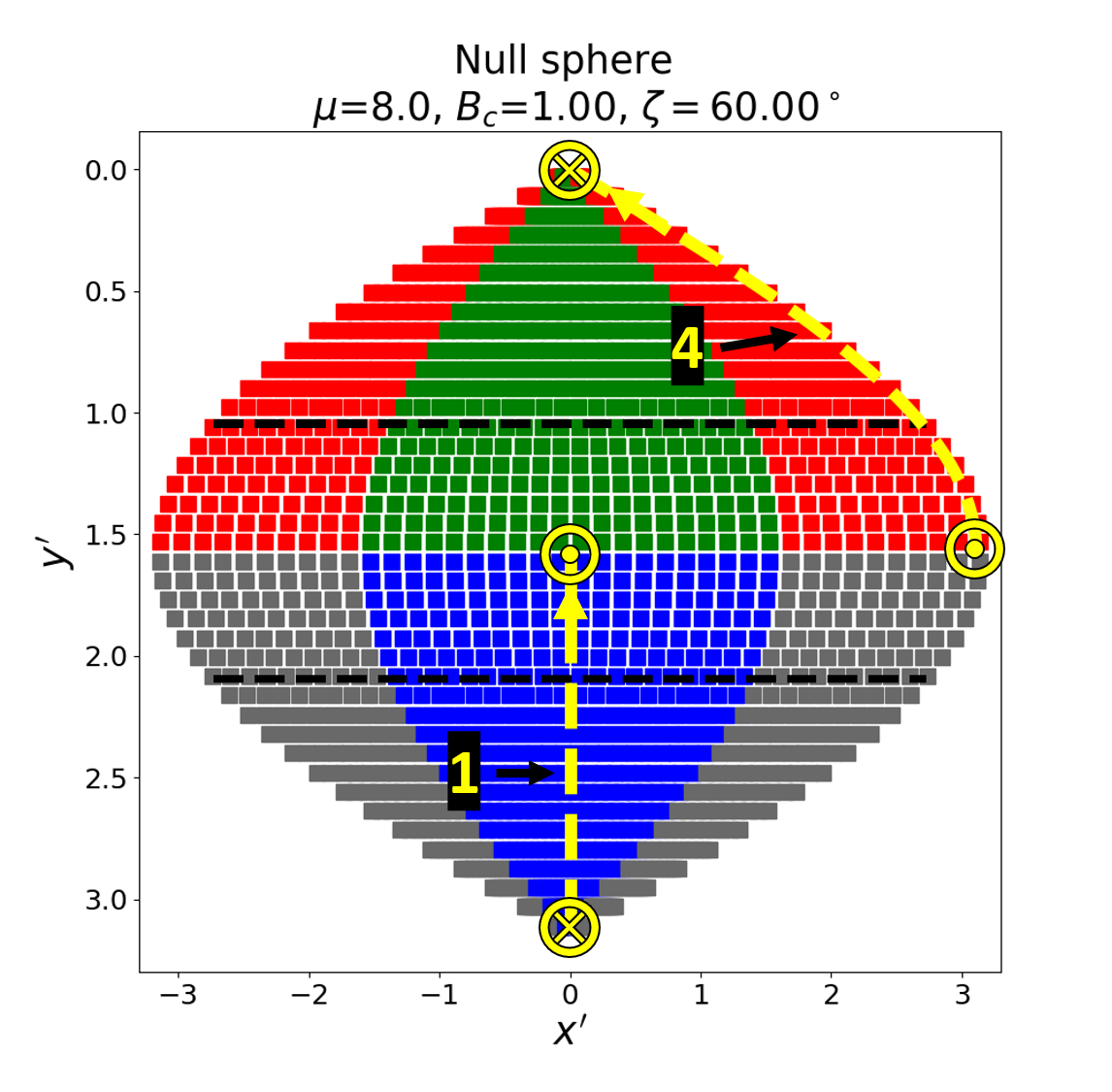}}
	\\
	\subfloat[][Far Sphere. Notice the two red regions are in fact continuous due to periodicity in $\phi$. Along the black lines, $B_r=0$ ]{\includegraphics[height=125pt]{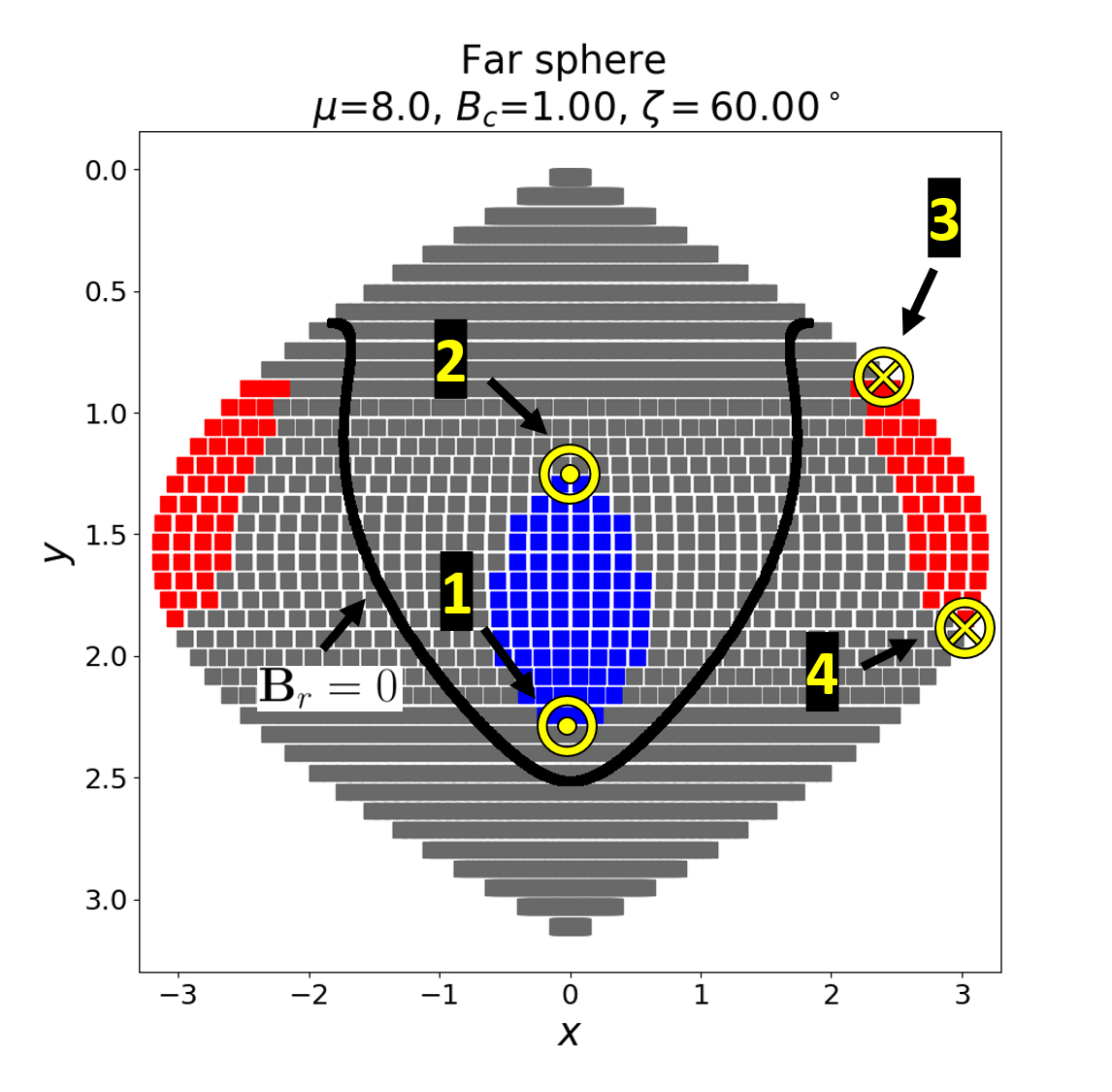}}
	\\
	\subfloat[][Dipole Sphere. Along the black dotted line, $B_r=0$]{\includegraphics[height=125pt]{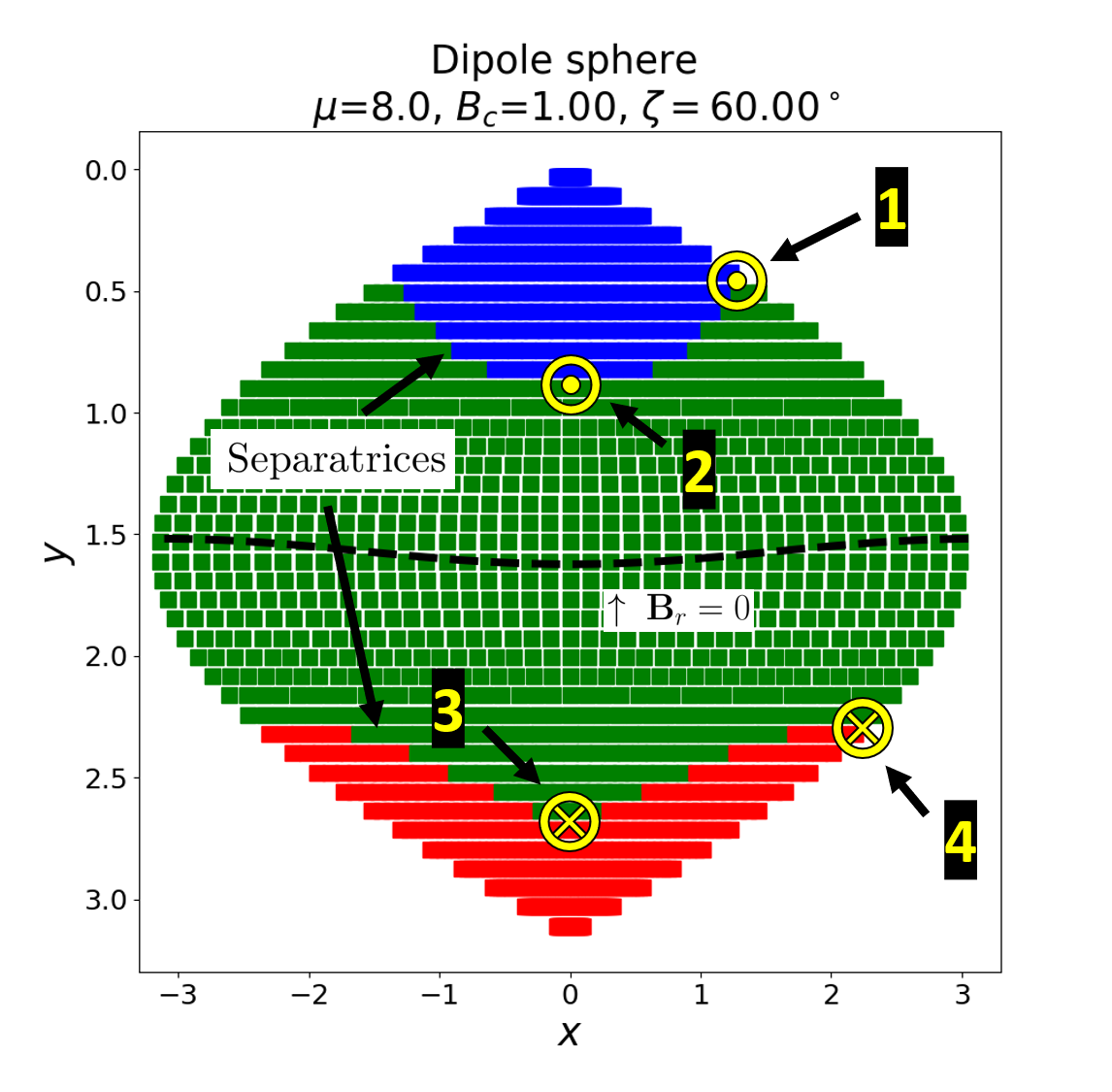}}
	\caption{The characteristic field lines 1 and 2 map from the dipole sphere to land on the null sphere and travel along the sphere at a constant azimuthal angle until leaving the sphere to travel to the far-sphere. The characteristic field lines 3 and 4 follow the same path idea, but in reverse.}
	\label{fig:4-spheres}
\end{figure}

In Fig. \ref{fig:4-spheres} we use the sinusoidal map projection, 
\begin{equation}
x' = \varphi\sin\theta, \quad y'=\theta
\label{eqn:Sin_map_projection}
\end{equation}
in order to eliminate the appearance of degenerate points on the sphere (i.e. $(\theta_1,\phi_1)=(\pi/2,2\pi)$ and $(\theta_2,\phi_2)=(\pi/3,2\pi)$).
The null sphere coordinate system $(x',y')$ has been rotated from cartesian coordinates $(x,y,z)$ such that the rotated spherical coordinate system $(\theta',\varphi')$ aligns the $z-axis$ along the spine away from the dipole sphere and has been projected into the sinusoidal map. 

\vspace{0.5cm}\textit{a. Dipole sphere: Fig. \ref{fig:4-spheres}a}\\\newline
The dipole sphere has field lines of three topological types (Fig. \ref{fig:4-spheres}a). Along the change in topological type as $\theta:0\to\pi$ lies two separatrices: the region between the blue and green types, or green and red types. A fundamental property of magnetic nulls is that two field-lines that lie arbitrarily close to each other at one point in space may end up arbitrarily far away from each other if they pass near a null. Equivalently, field-lines which approach the separatrix infinitesimally will pass closer and closer to one of the respective nulls. The dipole separatrices account for half of the flux of each null, with the other half coming from the far-sphere separatrices.

\vspace{0.5cm}\textit{b. Far sphere: Fig. \ref{fig:4-spheres}c}\\\newline
The far sphere also contains three topological types, however the surface-surface field-lines are now $\infty$-$\infty$ field-lines. The far-sphere separatrix additionally maps directly into the magnetic nulls. 

\vspace{0.5cm}\textit{c. Null spheres: Fig. \ref{fig:4-spheres}b}\\\newline
There are two null spheres in our model, however they differ only in magnetic field sign and are symmetric.

The null sphere acts as a separator of topological types: all four topological types are observed on a sphere placed about a null, Fig. \ref{fig:4-spheres}b. The black dashed lines denote the angles along which $B_r=0$ and provide a definition for the spine and fan regions of the null-sphere. The fan region is a ribbon of flux and so is centered about $\theta=\pi/2$. The spine region is a thin column of flux along a single pole through the sphere and so does not map out a continuous region along the null sphere but rather two distinct pieces. If $\theta_0$ is the angle at which $B_r=0$ $\forall\varphi$, then $|\theta-\pi/2|>\theta_0$ is a spine point and $|\theta-\pi/2|<\theta_0$ is a fan point. See Sec. \ref{sec:F-L_Properties-NullSphere} for more discussion.

\vspace{1cm}


The fan of a null is populated by field lines which lie along one separatrix for $y\neq0$. At $y=0$, that separatrix is populated by a field line from the other null's spine. The same interpretation is true for the other null, with the caveat that the field-line directions will all reverse for the other null. 


The null sphere is the intersection of surfaces dividing volumes in space composed of field-lines of single topological type. 

First we consider only far-sphere separatrix contributions to the null sphere. The separatrices of the far-sphere map out a closed surface on either side of which field-lines are either of $\infty-\infty$ type or $\infty$-Dipole sphere / Dipole sphere-$\infty$ type. The separatrix maps into either the spine or fan of the null, depending on which null is being considered. The flux contribution to this null is not the complete surface but rather a complete surface without a single line lying in the symmetry plane. This missing contribution of flux is attached to the other null; it is the spine of the other null.  


In conclusion, the null sphere is the meeting point between separator surfaces. The four characteristic field-lines divide these surfaces in the symmetry plane.

 The flux passing through a sphere placed about a null is fundamentally tied to the flux passing through the dipole separatrix. This will lead to a clear argument against the relevance of of magnetic null points themselves to reconnection processes.  Instead a large volume, defined by the distinguishability distance $\Delta_d$, that includes the null determines reconnection properties.

\subsection{Sphere about a null \label{sec:F-L_Properties-NullSphere}}
Here we discuss a null-characterizing structure: a sphere of radius $r$ placed about a null; see Fig. \ref{fig:Null-Sphere}. The flux through this sphere is invariant to the current $\b{j}_0$ whereas the spine-fan structure varies with $\b{j}_0$. $\b{j}_0$ is not invariant in an arbitrary ideal magnetic evolution; it may change to preserve topological properties of the field  \cite{Boozer:MR-with-null+X-points}\cite{Parnell-1996}. Restrictions on such an evolution may be found in \cite{Hornig_1996}. For more discussion on currents related to nulls, see Sec. \ref{sec:F-L_Properties-Current_at_null}. 

In the next section, we will see that this interpretation provides insight through a frozen-flux argument.

\subsubsection{Null sphere flux \label{sec:F-L_Properties-NullSphere_Flux}}

\begin{figure}[h!]
	\centering
	\vspace{0cm}
	\subfloat[][]{\includegraphics[height=150pt]{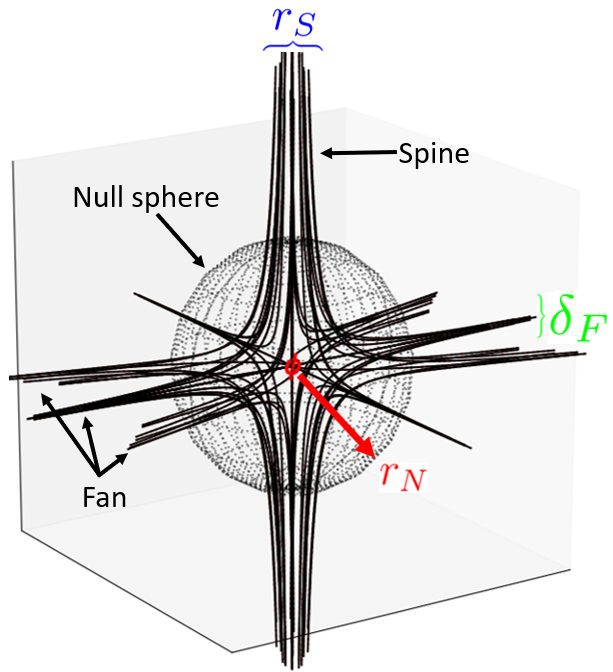}}
	\caption{Field-line structure near a magnetic null. Points near but not on the spine or fan trace out paths remniscent of the spine-fan structure albeit at a greater distance from the null. The quadrupole term $Q_n$ describes the field-line flow along the fan: as $Q_n\to 1$ the fan flux collapses to a spine-like structure, causing the overall spine-fan structure to resemble a cross.}
	\label{fig:Null-Sphere}
\end{figure}

The flux passing through a sphere about the null is the radial flux as measured from a null-centered frame. Asumming $\b{j}=0$ in the  representation for the near-null magnetic field (\ref{eqn:Null-field}), we have
\begin{equation}
\b{B}_{r} = r \frac{B_n}{2a} \big( (1 - 3\cos^2\theta) + Q_n \ss\theta(1-2\ss\varphi)\big)
\label{eqn:B_radial}
\end{equation}
the spine-fan separatrix may be found by solving $\b{B}_r=0$, yielding
\begin{equation}
\theta = \tan^{-1}\bigg( \sqrt{\dfrac{3}{1+Q_n(1-2\ss\varphi)}} \bigg)
\label{eqn:SF_separatrix}
\end{equation}

We now consider the flux passing through the null sphere. As we have shown previously, the flux through the null sphere is tied to the separatrices of the far sphere and the dipole sphere. For reasons which will be clarified later, we restrict our attention to the flux relationship between the dipole sphere and the null sphere. 

By placing a sphere of radius $a_N$ about the null (Fig.\ref{fig:Null-Sphere}) we may relate the flux from the separatrix to the flux at the null.  

It is useful to define spatial parameters and to assume the spine-fan structures have a finite size. The fan is a plane of flux passing into the null; call its width $\delta_F$. The spine is a cylindrical column of flux; call its radius $r_S$. Notice that  $\delta_F,r_S$ are not arbitrary length scales, rather they are set by the minimum length scale of distinguishability of the system. In this system, this occurs at the dipole separatrix.

Lets say field-lines pass into the null along the fan and out of the null along the spine. The flux passing through the surface of the sphere scales with $r$, eqn (\ref{eqn:B_radial}). We choose to suppress terms varying with $\theta$, $\phi$ as we are only interested in flux scaling.  That is, $B_{r}(a_N,\theta,\varphi)\sim r$. 

Thus, the flux passing into the sphere about the null from the fan is 
\begin{equation}
\Phi_F = B_r(a_N,\theta,\varphi) (2 \pi a_N \delta_F) \sim a_N^2 \delta_F
\label{eqn:Phi_F}
\end{equation}
The flux passing out of the null along the spine is 
\begin{equation}
\Phi_S = B_r(a_N,\theta,\varphi)(\pi r_S^2) \sim a_N r_S^2
\label{eqn:Phi_S}
\end{equation}

The flux contribution of a single topological type to the null sphere is vanishing; $\Phi_F=-\Phi_S$. Suppressing signs, we find $a_N r_S^2 \sim a_N^2 \delta_F$ or 
\begin{equation}
r_S \sim \sqrt{a_N \delta_F}
\label{eqn:NS_flux_eqn1}
\end{equation} 

We now consider the flux relation between the null fan of one topological type and the flux of the dipole separatrix. Necessarily, $\Phi_F = \Phi_{Sep}$ where $\Phi_{Sep}$ is the flux contribution from one dipole separatrix to the null fan. Remembering the radius of our dipole sphere is $a$, we have 
\begin{equation}
\Phi_{Sep} = B(\b{x}) 2 \pi a \delta_S \propto a^2 \delta_S
\label{eqn:Phi_Sep}
\end{equation}
where $\delta_S$ is the width of the dipole separatrix. On the dipole separatrix, $B(\b{x})\approx B_d(\b{x}) \propto a$.

Equating $\Phi_{Sep}=\Phi_S$, we have $2 \pi a \delta_S = 4 \pi a_N^2 B_r $. Since $B_r\sim r$ close to the null, choose $r=a$ so we have scaling as 
\begin{equation}
a_N = \sqrt[3]{\delta_s a^2}
\label{eqn:NS_flux_eqn2}
\end{equation} 
Notice that we are only talking about flux contributions from one dipole separatrix to one null fan, and subsequently one null spine.

Thus, the smallest distance in this system is $\delta_S$, the dipole separatrix width. In particular, choosing $a_N$ small enough such that the lowest order terms in its Taylor expansion dominates, we require a dipole separatrix of width $\delta_S = a_N^3/a^2$. 

We do not consider the relation between the null and far sphere fluxes here. The field strength in the far field will not impose a smaller relevant length scale  since the constant field strength is weaker than the field strength on the dipole separatrices.

\section{Field properties \label{sec:F-L_Properties}}
\subsection{Variation of null properties with $\zeta$ \label{sec:F-L_Properties-Zeta_variation}}
\subsubsection{Null location \label{sec:F-L_Properties-Null_loc_zeta}}
Nulls are points in space which satisfy $\b{B}_d=-\b{B}_c$. Nulls lie at a distance $a_n \sim \sqrt[3]{\mu/B_c}$, with the $\theta$ coordinate determined numerically. Along the $y=0$ plane, $\b{B}_{\varphi}=0$ and so  $\varphi=0$ for all nulls regardless of $\zeta$; see Sec \ref{sec:F-L_Properties-Symmetry_plane}. 

If $\mu/a^3<B_c$, the nulls lie within the dipole sphere which has radius $a=a_D$. This is not relevant to the discussion of nulls given in this paper. In this paper, $\mu/a^3>B_c$ strictly.

At $\zeta=0$, the nulls lie at the poles of the dipole sphere: $\theta=0$,$\pi$. At $\zeta=\pi$, a line null is formed about the equator of the dipole sphere. The line null in coordinate form may be expressed $\theta=\pi/2$, $\varphi\in[0,2\pi]$. Any magnetic perturbation away from the $\zeta=\pi$ configuration collapses the line null into two point nulls, a well-known topological property of nulls. For any angle $0<\zeta<\pi$, the null continuously travels between these two extremes along the $y=0$ plane.

\subsubsection{Spine-fan variation with $\zeta$ \label{sec:F-L_Properties-SF_zeta}}
\begin{figure}[h!]
	\centering
	\vspace{0cm}
	\subfloat[][Spine-fan for $Q_n=0$. The spine-fan takes on its usual form: a column of field lines form the spine and a plane of field-lines form the fan]{\includegraphics[height=130pt]{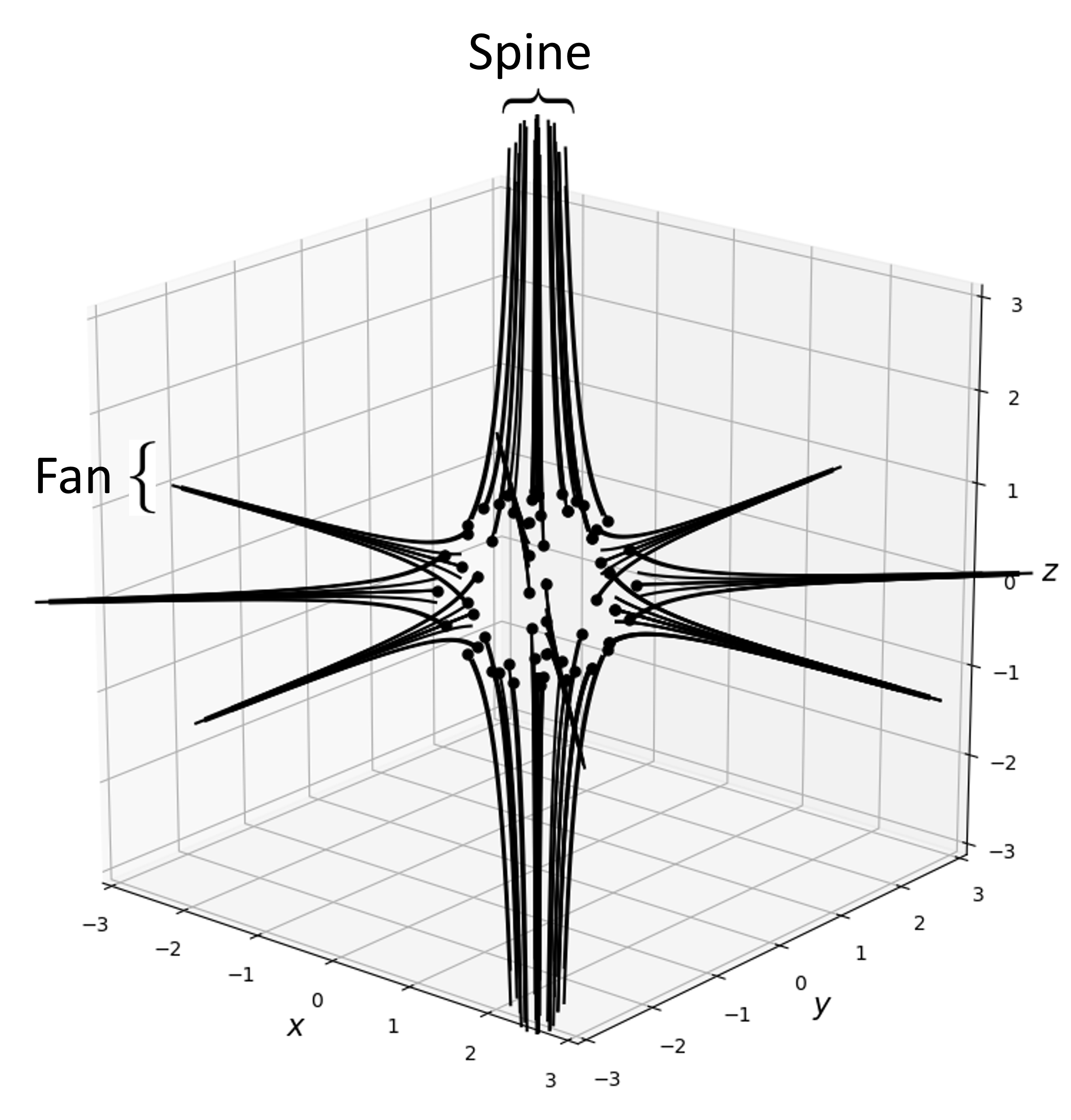}}
	\subfloat[][Spine-fan for $Q_n=1$. Here the spine widens into a constant-radius column of field lines while the fan collapses into another column perpendicular to the spine]{\includegraphics[height=130pt]{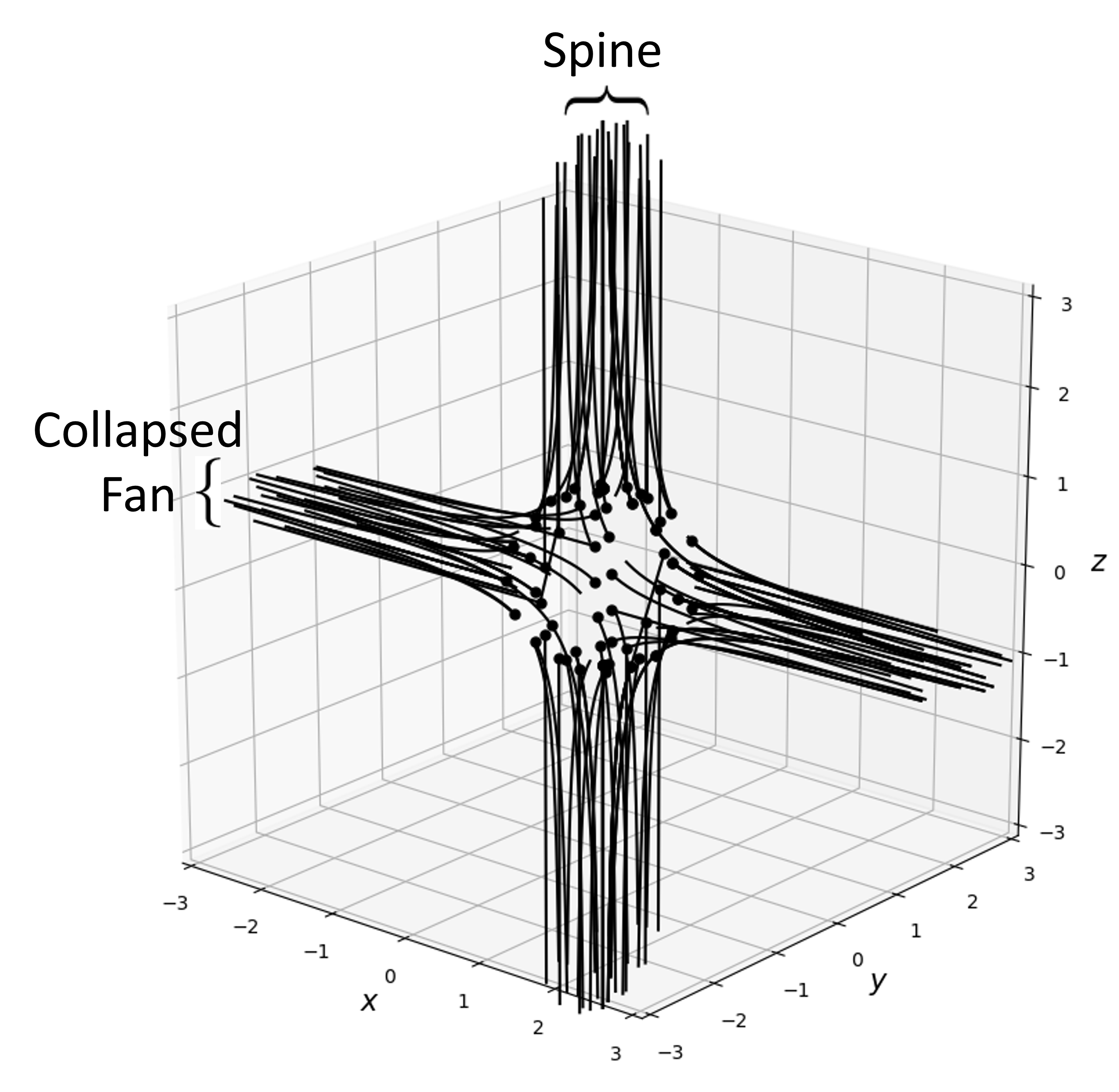}}
	\caption{}
	\label{fig:SF_limiting_cases}
\end{figure}

\begin{figure}[h!]
	\centering
	\vspace{0cm}
	\subfloat[][]{\includegraphics[height=150pt]{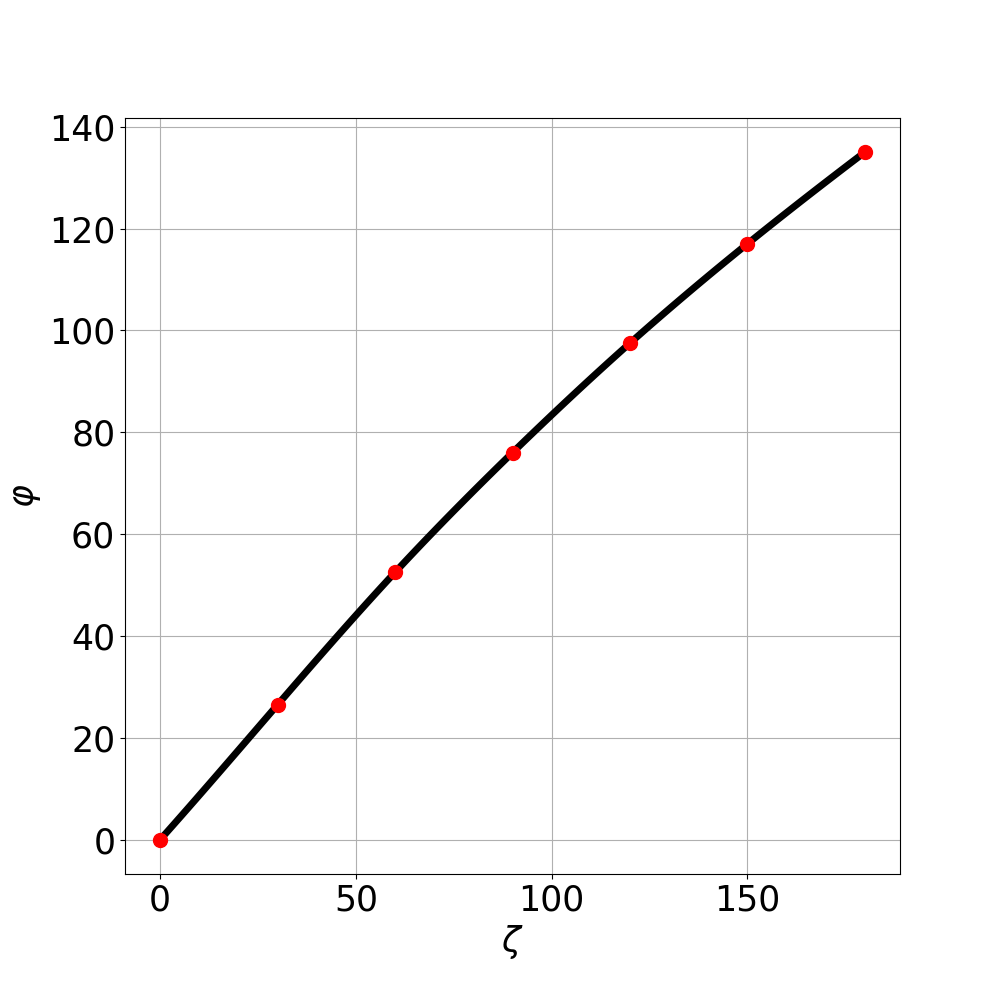}}
	\\
	\subfloat[][]{\includegraphics[height=150pt]{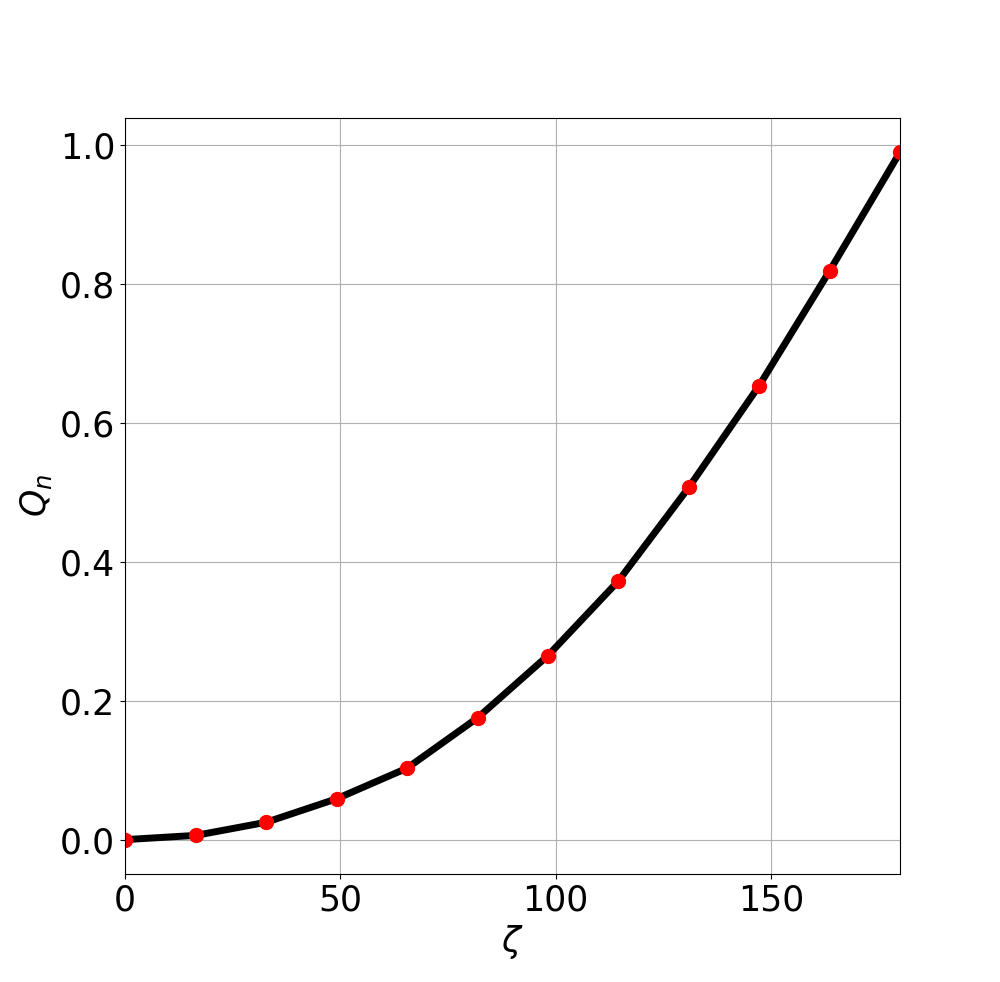}}
	\caption{(a) Rotation of spine orientation with $\zeta$. $\varphi$ is measured with respect to $\h{z}$, (b) Quadrupole term $Q_{n}(\zeta)$ which describes different topologies of the spine-fan structure}
	\label{fig:SF_variation_zeta}
\end{figure}

The spine-fan structure (eqn. (\ref{eqn:Null-field})) varies with the background magnetic field angle $\zeta$. Specifically, $Q_n=Q_n(\zeta)$ and the spine-fan structure rotates with $\zeta$; see Figs. \ref{fig:SF_limiting_cases},\ref{fig:SF_variation_zeta}. 
$Q_n$ was calculated here using least squares. The topological change occurs with $Q_n$ and may be described by the two limiting cases: $\zeta=0$ and $\zeta=\pi$. For $\zeta=0$, $Q_n=0$ and so the spine-fan structure is similar to that pictured in Fig. \ref{fig:Null-Sphere}; the fan is a plane of field-lines and the spine is a column of field-lines. As $\zeta\to\pi$, $Q_n \to 1$ and the fan becomes strongly unidirectional along one axis, collapsing the fan into a spine-like structure. At $\zeta=\pi$ the spine-fan structure converts into a spine-spine structure, two poles of magnetic field  lines, while simultaneously a line null forms along the dipole sphere's equator; the spine-fan collapses, forming a 2D null.

\subsection{Field-line distinguishability \label{sec:F-L_Properties_Distinguishability}}
Magnetic reconnection is fundamentally a loss of field-line distinguishability. Ideal evolutions assume field-line resolution is perfect, or a breaking of the frozen-in condition. This breakdown may come from any physical effect limiting field line distinguishability over a length scale $\Delta_d$, such as resistivity. From Boozer \cite{Boozer:MR-with-null+X-points} we know a universal physical effect, electron inertia, provides the minimal limit of field-line distinguishability, the electron skin depth $c/\omega_{pe}$. We may typically say that $\Delta_d>c/\omega_{pe}$; the electron skin depth breaks the frozen-in condition even when resistivity is neglected.

%

Boozer finds the field-line evolution equation with electron inertia effects is
\begin{equation}
\dfrac{\p}{\p t}\bigg( \b{B} - \nabla\times\Big(\nabla\times\bigg(\frac{c}{\omega_{pe}}\b{B}\bigg) \bigg) = \nabla\times\big(\b{u}\times\b{B}\big)
\label{eqn:Mag-evol-eqn}
\end{equation}
where $c/\omega_{pe}$ is the well-known electron skin depth. $\mathbf{u}$ here is the magnetic field-line velocity; using $\mathbf{u}$ instead of the plasma flow velocity simplifies the discussion. If the electron skin depth is constant, the equation reduces to 
\begin{equation}
\dfrac{\p}{\p t}\bigg( \b{B} - \Big(\frac{c}{\omega_{pe}}\Big)^2 \nabla^2 \b{B}
 \bigg) = \nabla\times\big(\b{u}\times\b{B}\big)
\notag
\end{equation}

Thus, in any evolving magnetic field with spatial variation, the limit of field-line distinguishability is the electron skin depth, $c/\omega_{pe}$. Again we stress that the field must be evolving; without any temporal variation in the field there is no loss of field-line identity. Only in an evolution does field-line topology preservation have meaning.

\subsubsection{Application to nulls in interacting dipolar fields \label{sec:F-L_Properties-Distinguishablity_DipolarNulls}}
We have shown that Ohm's law with electron inertia effects, a statement that the electron mass is finite, reveals a fundamental limit of field-line distinguishability; the electron skin depth. 

Further, we have shown that the smallest length scale in our system is the dipole separatrix width $\delta_S$ which has the relation
\begin{equation}
\sqrt[3]{a^2 \delta_{S}}>>a_N
\notag
\end{equation}

Combined, we may consider the case where $\delta_S = c/\omega_{pe}$. 
During an evolution field lines that lie within $\Delta_d$ of a separatrix are indistinguishable and will reconnect in any evolution.
Thus, $\sqrt[3]{a^2 (c/\omega_{pe})}>>a_N$ and we may state the the global spine-fan structure of magnetic nulls is enveloped by a surface below which all field-lines are indistinguishable in an evolution. 

More generally, all separatrices in this system become fuzzy below the electron skin depth.

\section{Constraints on the current}
\subsection{Current density at a null \label{sec:F-L_Properties-Current_at_null} }

The general expression for the magnetic field near a null, Equation (1), includes the current density $\mathbf{j}_0$ at the null.  A remarkable feature of this general form is that it requires that $\mathbf{j}_0=0$ unless $Q_n=\pm1$ or the plasma exerts a force with a non-zero curl at the null. $Q_n$ can always be made to lie in $[0,1]$ by choice of coordinates about the null.

The sum of the force exerted by the plasma and the Lorentz force $\mathbf{f}_L \equiv \mathbf{j}\times\mathbf{B}$ exerted by the magnetic field must vanish everywhere.  Since both $\mathbf{j}$ and $\mathbf{B}$ are divergence free, a vector identity implies the curl of the Lorentz force $\mathbf{\nabla}\times\mathbf{f}_L=\mathbf{B}\cdot\mathbf{\nabla}\mathbf{j} - \mathbf{j}\cdot\mathbf{\nabla}\mathbf{B}$.  This general expression coupled with  Equation (1) for $\mathbf{B}$ near the null requires that at the null
\begin{equation}
\mathbf{\nabla}\times \mathbf{f}_L = -\mathbf{j}_0\cdot\tensor{M},
\end{equation}
where $\tensor{M}$ is the matrix given in Equation (2).  The curl of the Lorentz force vanishes only if $(1+Q_n)\mathbf{j}_0\cdot\hat{x}=0$, $(1-Q_n)\mathbf{j}_0\cdot\hat{y}=0$, and $\mathbf{j}_0\cdot\hat{z}=0$.  

If $Q_n\in[0,1]$ by choice of coordinates then the only way to obtain a curl-free force is to have $Q_n=1$ and the current $\vec{j}_0$ in the $\hat{y}$ direction.  Fig. \ref{fig:SF_limiting_cases}b shows the $\hat{y}$ direction when $Q_n=1$.

The force from scalar plasma pressure is curl free; only the curl of the plasma inertial-force, $\rho (\partial\mathbf{v}/\partial t+\mathbf{v}\cdot\mathbf{\nabla}\mathbf{v})$, is always available to balance $\mathbf{\nabla}\times \mathbf{f}_L$.

\subsection{Development of a current along $\mathbf{B}$ \label{sec:F-L_Properties-Alfven_Transit_time}}

The magnetic field lines that pass through the two separaticies of the dipole sphere pass through the null and represent locations of topology changes in the magnetic field lines.  Field lines that lie closer to the separatrix than $\Delta_d \gtrsim c/\omega_{pe}$ are indistinguishable and they spread apart as they approach a nulls.  This and the constraint on the current density at the null, Section \ref{sec:F-L_Properties-Current_at_null}, makes an extremely large current density improbable at the separatrix.

At important set of magnetic field lines exit the dipole sphere close to one separatrix and reenter close to the other but pass far enough away from the separaticies to be distinguishable.  In the limit as plasma resistivity goes to zero, the topology-conserving properties of the ideal magnetic evolution equation should cause large current densities to be produced along these magnetic field lines if the strength or the orientation of the spatially uniform field is changed.  A sufficiently long time after the change is made these current densities should be essentially constant along the magnetic field lines.  The reason is that the $\mathbf{j}$ is divergence free, which can be written as 
\begin{equation}
\mathbf{B}\cdot\mathbf{\nabla}\frac{j_{||}}{B} = \mathbf{B}\cdot\mathbf{\nabla}\times \left(\frac{\mathbf{f}_L}{B^2}\right),
\end{equation}
so any variation of a singular $j_{||}/B$ along a field line must be balanced by a singular Lorentz force.  The natural time for $j_{||}/B$ to relax to a constant value is given by the balance between the Lorentz force and the plasma inertial force, which implies the relaxation requires a transit time of the shear-Alfv\'en wave \cite{Boozer-Elder}.

When a change in the external field is made that would produce a topology change of field lines that pass near the separatrix, a shear Alfv\'en wave must transit from one intersection of the perturbed field line with the dipole sphere to the other for the appropriate $j_{||}/B$ current to be established to conserve topology.  This is analogous to the proof of Hahm and Kulsrud in Section III of of their 1985 paper \cite{Hahm-Kulsrud} that in an ideal evolution the current density increases only linearly in time near a resonantly perturbed rational surface although current density must become singular in the infinite time limit \cite{Boozer-Pomphrey}.   A linear increase in $j_{||}$ in time due to this Alfv\'en relaxation effect is too slow to explain the fast breaking of magnetic surfaces observed during tokamak disruptions through a reconnection directly caused by $\eta j_{||}$.  A linear increase of the current density in time was also found in \cite{Boozer-Elder} for a smoothly driven ideal evolution. 

The time can be estimated for a shielding current to flow in a sufficiently narrow channel $\delta$ near the separatrices on the dipole sphere that resistive diffusion becomes important.  The magnetic Reynolds number $R_m$ is the ratio of the time scale for resistive diffusion to the time scale for the ideal evolution.  Let $j_c\equiv B_{avg}/(\mu_0 a)$ be a characteristic current density and with $B_{avg}$ defined in the caption to Figure 8.  Resistive diffusion competes with the ideal evolution when the evolving current density reaches $j\sim j_c R_m$.  The current density required to maintain the magnetic field line connections by flowing within a distance $\delta$ of the separatrix on the dipole sphere is $j = j_c a/\delta$.  This is in analogy to the result of Hahm and Kulsrud \cite{Hahm-Kulsrud} for a resonantly perturbed rational surface.  As will be shown, an Alfv\'en transit time, $\tau_A\propto 1/\delta^P$ is required for a parallel current to develop within a distance $\delta$ of a separatrix, where $P$ is a power given in Figure \ref{eqn:Alfven_transit_time}a. Consequently, the time required for a sufficiently intense current density to arise for resistive diffusion to compete with the ideal evolution in the neighborhood of the separatrix is $\sim \tau_c R_m^{1/P}$, where the characteristic Alfv\'en time is $\tau_c \equiv a/B_{avg}$.  When $R_m>>1$ this time can be too long for direct resistive diffusion to compete with $c/\omega_{pe}$ or with the exponential enhancement of the effect or resistivity on reconnection, which is illustrated in \cite{Boozer-Elder}.

Fig. 8a illustrates the behavior of the Alfv\'en transit time $\tau_A$ along magnetic field lines that intercept the dipole sphere within a distance $\delta$ of a separatrix,
\begin{equation}
\tau_A = \int \dfrac{ds}{|B(\mbox{x})|} \sim \dfrac{c_0}{\delta^P}
\label{eqn:Alfven_transit_time}
\end{equation}
This is the definition of the Alvf\'en transit time assuming a constant density.
$\tau_A$ exhibits power-law dependence on distance from the separatrix with the strength of the power law depending on $\varphi$ and the orientation of the uniform background field $\zeta$. When  P=1 the separation from the separatrix, $\delta$, $\tau_A$ scales as $\tau_A \propto 1/\delta$, which is the scaling found by Hahm and Kulsrud as a resonantly perturbed rational surface is approached.  In general, $\tau_A \propto 1/ \delta^P$.

\begin{figure}[h!]
\centering
\vspace{0cm}
\hspace*{-0.5cm}
\subfloat[][]{\includegraphics[height=120pt]{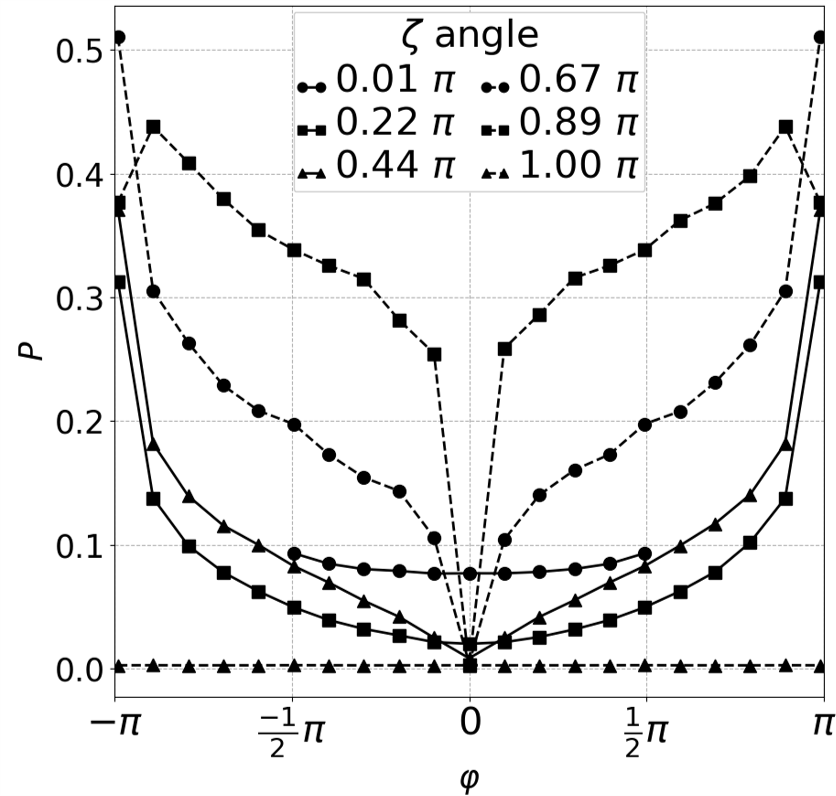}}
\subfloat[][]{\includegraphics[height=120pt]{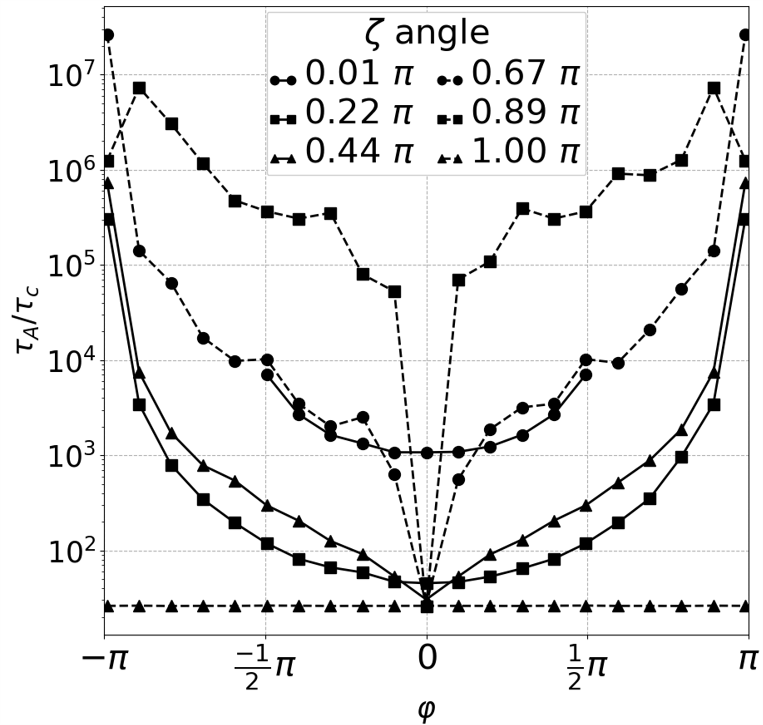}}
\caption{ (a) $P(\varphi,\zeta)$ and (b) $\tau_A/\tau_C$ where $\tau_A$ was computed for $\delta/a=10^{-15}$ and $\tau_C$ is a characteristic Alfv\'en transit time defined as the ratio of the dipole radius $a$ to the average field strength on the dipole sphere for $\zeta=0$,  $\tau_C = a/B_{avg} = 4a/(2\mu/a^3 + B_c)$ where $B_{avg} = \langle B \rangle \big|_{r=a}=\oint \mathbf{B}\cdot d\mathbf{a}/4\pi a^2$ on the dipole sphere surface when $\zeta=0$.}
\label{fig:Alfven_transit_time}
\end{figure}

Fig. \ref{fig:Alfven_transit_time}b is the Alfv\'en transit time for each point launched $\delta=10^{-15}$ away from the dipole separatrix, in other words the closest possible to the dipole separatrix that numerical accuracy allows.

On the symmetry plane ($\varphi=0,\pm\pi$), $P$ and $\tau_A$ reach a minimum value. Notice for $\zeta=\pi$, every $\varphi$ angle is a symmetry plane and so the $P$, $\tau_A$ values converge to their values otherwise found on the symmetry plane. However, for $\zeta\to 0$ the dipole separatrix is closer to the poles of the dipole and consequently pass near to the opposite null. Thus $P$ and $\tau_A$ are greater than their typical values when $\zeta\to 0$. The behavior on the symmetry plane is an open research question here. One final caveat is that the fit to the $\delta$ dependence may not be a power law on the symmetry plane, but instead of the form $\tau_A^{Symm.}\sim c_0 \ln(c_1 \delta/a) + c_2$; this may be anticipated by considering eqn. \ref{eqn:Alfven_transit_time} analytically where $\mathbf{B}$ is linear about the null and an arbitrary axises rotation allows for $ds\to dx$ and $\mathbf{B}(\mathbf{X})\to\mathbf{B}(x)$.

To obtain the power law dependence of $\tau_A$, values of $\delta/a$ between $10^{-15}$ and $10^{-1}$ were used.

A comment on the effect of numerical accuracy in this calculation: field-line trajectories are evaluated by integrating $d\mathbf{x}/dl = \mathbf{B}(\mathbf{X})$ where $\mathbf{X}$ is a position in space. Both near the null, and near remnants of the line null which forms when $\zeta=\pi$ have vanishing and near-vanishing field strength, respectively. In these regions, $|\mathbf{B}|\to 0$ and as a result is subject to numerical inaccuracy. This is further complicated in that the quantities used to evaluate $\tau_A$ as we are again dealing with quantities not much greater than numerical accuracy. As a result, integration of $\tau_A$ is subtle.

\color{black}

\section{Discussion \label{sec:Summary}}
We have reproduced the classic null-generating model of Cowley with additional considerations of (i) mappings between four characteristic spheres of the system, (ii) separatrices of the system, (iii) flux considerations through a sphere placed about a null and (iv) the fundamental and ubiquitous effect on field-line distinguishability of electron inertia. Additionally we have documented the change of the spine-fan variation with uniform background field angle $\zeta$. 

By considering the system as a mapping between the dipole sphere, null spheres, and far-sphere combined with the concept of topological type fundamental to magnetic nulls, we find the separatrices of each sphere are tied to each of the others through the null sphere. Flux conservation relates the areas through which flux passes at the dipole and far-sphere separatrices to the flux passing through a sphere of radius $a_N$ placed about a null. In concert with the limit of field line distinguishability, the electron skin depth $c/\omega_{pe}$, we may conclude that the areal constraint at the dipole and far-sphere separatrices implies that magnetic nulls and their spine-fan structures are globally enclosed by a volume within which magnetic field-lines are indistinguishable. Thus, we conclude that the importance of magnetic nulls is in (i) its ability to bring field-lines of different topological type close, allowing field-lines of different boundary conditions to pass near one another and (ii) the spine-fan structure of magnetic nulls which through (i) can lead to separator surfaces which support large currents. The width of the current channel is affected by the $(a^2c/\omega_p)^{1/3}$ limit on distinguishability.

However, since electrons are the lightest current carrier, $c/\omega_{pe}$ effects are always present. This effect has a relaxing affect on magnetic topologies and current singularities. 

In an ideal theory, a point null can be approached infinitesimally and provide singular current densities. If instead we account for field-line distinguishability due to electron skin depth, there exists a volume of field-line indistinguishablitity which envelops both the null and its global spine-fan structure. Field-lines do not retain their connections within this surface. Electron inertia effects smooth out any sharpness in the magnetic field which would yield singular currents as well as renders obsolete the loss of field-line identity at the point where $\mathbf{B}=0$. Instead, it is the magnetic structures formed to satisfy $\nabla\cdot\mathbf{B}=0$ at the point where $\mathbf{B}=0$ which make the null relevant. 

Finite spatial resolution in a numerical simulation has a similar effect as $\Delta_d$ on field-line distinguishability.  Indeed, Pariat et al [17] obtained magnetic reconnection in an otherwise ideal simulation due to the finite grid.  Olshevsky et al [16] included electron inertia in their simulations of multi-null reconnections and found the narrowest current channels were broader than $c/\omega_{pe}$. 

In an ideal model field-lines will only be distinguishable up to numerical accuracy. Numerical innaccuracies provide a mechanism for reconnection just as $\Delta_d$ limits field-line distinguishability. On any shorter length-scale, field line identity is lost. 

%

The effect of a quadrupole term has been considered but not included in this work. The higher-order term breaks the magnetic surface, which is the $y=0$ plane, but otherwise does not change the fundamental properties of the system when the quadrupole term is too weak compared to the dipole term to produce additional nulls outside of the dipole sphere. This is a topic of continuing study.

Another consideration is the effect of the dipole sphere and far sphere's boundary conditions. For example, a perfectly conducting boundary condition would allow current to flow along field-lines attached to the dipole sphere. If the boundary condition were insulating, a steady current could not flow along the field-lines and the topology of the field lines would break on an Alfv\'enic time scale \cite{Boozer-MR_in_space}.

When the dipole sphere is a perfect conductor surrounded by a plasma with a sufficiently low resistivity $\eta$ that the magnetic Reynolds number is enormous for a characteristic evolution time, $R_m>>1$, it takes a very long time for shielding currents to concentrate sufficiently close to field lines that pass near the dipole-sphere separatrices for direct resistive diffusion to be relevant.  As $R_m$ goes to infinity, non-ideal effects of electron inertia on the $c/\omega_{pe}$ scale and of exponential enhancement of resistivity, as illustrated in\cite{Boozer-Elder}, dominate.

\section*{Acknowledgements}
This work was supported by the U.S. Department of Energy, Office of Science, Office of Fusion Energy Sciences under Award Numbers DE-FG02-95ER54333, DE-FG02-03ER54696, DE-SC0018424, and DE-SC0019479.

\section*{Data availability}
The data that support the findings of this study are available from the corresponding author
upon reasonable request

\end{document}